\documentclass[aip,rsi,reprint,floatfix]{revtex4-1}

\usepackage{xcolor}
\definecolor{xlinkcolor}{cmyk}{1,1,0,0}

\usepackage{microtype}
\usepackage[colorlinks, pdfusetitle, allcolors=xlinkcolor]{hyperref}
\usepackage[add-decimal-zero=false]{siunitx}
\usepackage{graphicx}
\usepackage{booktabs}
\usepackage[capitalise]{cleveref}

\renewcommand{\doibase}[0]{https://doi.org/}

\hypersetup{pdfauthor={Matthew Petroff, et al.}}

\begin{document}

\title{A 3D-printed broadband millimeter wave absorber}

\author{\href{https://orcid.org/0000-0002-4436-4215}{Matthew Petroff}}
\email{petroff@jhu.edu}
\author{\href{https://orcid.org/0000-0002-8412-630X}{John Appel}}
\affiliation{\mbox{Department of Physics \& Astronomy, Johns Hopkins University, Baltimore, Maryland 21218, USA}}
\author{\href{https://orcid.org/0000-0003-4189-0700}{Karwan Rostem}}
\affiliation{\mbox{Code 665, NASA Goddard Space Flight Center, Greenbelt, Maryland 20771, USA}}
\author{\href{https://orcid.org/0000-0001-8839-7206}{Charles L. Bennett}}
\author{\href{https://orcid.org/0000-0001-6976-180X}{Joseph Eimer}}
\author{\href{https://orcid.org/0000-0003-4496-6520}{Tobias Marriage}}
\affiliation{\mbox{Department of Physics \& Astronomy, Johns Hopkins University, Baltimore, Maryland 21218, USA}}
\author{Joshua Ramirez}
\affiliation{Johns Hopkins Applied Physics Laboratory, Laurel, Maryland 20723, USA}
\author{\href{https://orcid.org/0000-0002-7567-4451}{Edward J. Wollack}}
\affiliation{\mbox{Code 665, NASA Goddard Space Flight Center, Greenbelt, Maryland 20771, USA}}

\begin{abstract}
We present the design, manufacturing technique, and characterization of a 3D-printed broadband graded index millimeter wave absorber. The absorber is additively manufactured using a fused filament fabrication (FFF) 3D printer out of a carbon-loaded high impact polystyrene (HIPS) filament and is designed using a space-filling curve to optimize manufacturability using said process. The absorber's reflectivity is measured from \SIrange{63}{115}{\giga\hertz} and from \SIrange{140}{215}{\giga\hertz} and is compared to electromagnetic simulations. The intended application is for terminating stray light in Cosmic Microwave Background (CMB) telescopes, and the absorber has been shown to survive cryogenic thermal cycling.
\end{abstract}

\maketitle

\section{Introduction}

The ongoing development of increasingly sensitive cosmic microwave background (CMB) telescopes requires commensurate improvements in the control of systematic errors. One such source of error is from stray light, which needs to be controlled and terminated through the use of millimeter wave absorbers, both under ambient conditions and within cryogenic receivers.\cite{eimer2012,harrington2016} Since the field is moving towards the use of multichroic detectors with wide frequency bands in a shared optical path,\cite{s4tech} broadband absorbers are required. Per this application, the frequency range from \SIrange{30}{230}{\giga\hertz} is of interest, since it covers the CMB emission peak as well as synchrotron and thermal dust foregrounds. The maximum allowable reflectivity is dependent on where in the instrument the absorbers are used and is often difficult to quantify precisely; in general, lower reflectivity is better, but this must be considered in the context of cost, volume, and thermal design constraints. Broadband millimeter wave absorbers should also prove useful in W band radar applications or, more generically, as terminations or glint reduction media in an optical bench.

In recent years, additive manufacturing in the form of 3D printing has become increasingly common, in particular fused filament fabrication (FFF). [This is also referred to by the phrase ``fused deposition modeling'' and trademarked acronym FDM.] FFF-based printing works by extruding a plastic filament through a heated nozzle mounted on a CNC stage such that an object is built up layer-by-layer.\cite{jones2011} This allows for rapid prototyping and allows for manufacturing easily customized designs and one-off parts. In this work, this ease of customization is applied to the fabrication of millimeter wave absorbers. 3D printing has been previously used for manufacturing free space electromagnetic absorbers.\cite{sanz-izquierdo2014, kronberger2016,ren2018} To the best of our knowledge, however, only comparatively narrowband, resonator-based absorbers have been previously produced with such techniques.

Graded index absorbers generally take the form of an array of pyramidal structures.\cite{emerson1973} In the limit where the wavelength is similar to or greater than the feature size, the pyramids form a smooth gradient in effective permittivity, greatly reducing reflections. In the limit where the wavelength is smaller than the feature size, the pyramids cause the incident light to reflect off the absorber structure multiple times, with some radiation absorbed at each interaction. Since these two limits are governed by the feature pitch and the wavelength, it can be helpful to use the parameterization $p/\lambda_g$, where $p$ is the pitch and $\lambda_g = \lambda_0/\sqrt{\epsilon_r'}$ is the wavelength in the absorber material as a function of the free space wavelength, $\lambda_0$, and the real component of the absorber material's dielectric function, $\epsilon_r'$; this parameterization can be derived from the antenna theory analogue of a graded index absorber.\cite{chuss2017}

While a periodic pyramidal structure makes for an effective absorber, it is not well suited for FFF-based printing. When sliced into layers for FFF-based printing, each pyramid slice is disconnected from the others in a given layer. Thus, the filament extrusion process must be stopped and the filament retracted for each and every pyramid slice; this procedure prevents the creation of sharp points, since small amounts of plastic can be drawn back into the nozzle, and possibly causes stringing---thin strands of plastic stretched over gaps---that must be manually removed after the print is finished. Furthermore, prints are generally weakest along their layer lines, making points liable to break off. To avoid these issues, a geometric approximation of a space-filling curve is used, which fills the plane with a continuous wedge.

In this paper, we demonstrate that 3D printing allows for the rapid production of broadband millimeter wave absorbers. These easily customizable absorbers can achieve adequate performance for stray light termination at low cost and as a thermal plastic---not a foam---they can be used cryogenically. The fabricated absorbers are not intended to serve as extremely low reflectivity calibration targets, such as pyramidal array\cite{bai2017} and cone-shaped\cite{houtz2017} targets that have been detailed in the literature. Furthermore, due to the intended use case, only a low power regime has been evaluated. A comparison between the fabricated absorbers and other common types of low-profile graded index absorbers is presented in \cref{tab:comparison}.

\begin{table}
\centering
\begin{tabular}{@{\extracolsep{1em}}lccc@{}}
\toprule
& \multicolumn{3}{c}{Absorber type} \\
\cmidrule{2-4}
Property & Cast & Foam & 3D-printed \\
\midrule
Cryogenic compatibility & yes & no & yes \\
Easily customized & no & no & yes \\
Reflectivity & very low & low & low \\
\bottomrule
\end{tabular}
\caption{Comparison between types of low-profile graded index absorbers, specifically cast (or injection molded) pyramid arrays, foams with conductivity gradients, and the 3D-printed absorber presented in this work.}
\label{tab:comparison}
\end{table}

\section{Space-filling curves}

A space-filling curve is a curve that passes through every point of a two-dimensional region with positive Jordan measure, in this case area.\cite{sagan1994} [For such a curve, the Lebesgue covering (topological) dimension is equal to the Hausdorff (fractal) dimension. Thus, a space-filling curve is \emph{not} a fractal in the purest sense, since a fractal requires the Hausdorff dimension to be strictly greater than the geometric dimension.\cite{sagan1994,mandelbrot1977}] This type of mathematical monster was first described by Peano.\cite{peano1890} Shortly thereafter, Hilbert described another such curve but also described an iterative sequence of geometric approximations to his curve.\cite{hilbert1891} Importantly, this type of geometric approximation can be physically realized.

Space-filling curves are self-similar.\cite{sagan1994} Thus, absorbers created from different order geometric approximations of space-fillings curves will exhibit similar behavior. In this work, a Hilbert curve, which fills a square, is primarily used. However, other shapes are possible. For example, an absorber designed around a generalized Gosper curve,\cite{gardner1976,fukuda2001} which roughly fills a regular hexagon, was also prototyped, since it allows one to more easily cover a circle with absorber tiles. For $n>2$, all electromagnetic absorbers with $n$-fold rotational symmetry have no net polarization response.\cite{mackay1989} Although a Hilbert curve has only 1-fold symmetry---or 2-fold symmetry in the case of the loop variant, the Moore curve---geometric approximations of the Hilbert curve have asymptotically equal numbers of uniformly distributed horizontal and vertical line segments;\cite{moon2001} thus, they should not show a net polarization response in reflectivity, unlike 2-fold symmetric absorbers created from sets of parallel wedges.

To create an absorber, a wedge was modeled such that the peak of the wedge follows the centerline of a geometric approximation of the Hilbert curve. The wedge then extends down such that the halfway point between segments of the curve form troughs. Furthermore, only a shell is printed, leaving the inside hollow, to reduce printing time and save on material use. A rendering of a second order Hilbert curve absorber model can be seen in \cref{fig:simulation-model}.

A similar procedure can also be followed to create absorbers from other space-filling curves with appropriate symmetry, although this still only allows for a few different shapes. To create custom-shaped absorbers, a solid geometry intersection operation can be performed between the existing, square Hilbert curve absorber model and an extrusion of the desired absorber shape, e.g. a circular absorber can be created by computing the intersection between a cylinder and a Hilbert curve absorber model. By scaling the model's dimensions, it is easy to adjust the size and pitch of the absorber's features.

\section{Material selection}

Two different material candidates were tested, a commercially available carbon-loaded conductive polylactic acid (PLA) filament\cite{conductivepla} and a high impact polystyrene (HIPS) filament custom extruded\cite{filabot} from a commercially available carbon-loaded conductive HIPS pellet.\cite{conductivehips} The PLA has the advantage of being an off-the-shelf filament, and PLA is generally easier to print than HIPS. However, since the carbon-loaded HIPS comes in pellet form, it can easily be mixed with normal HIPS pellets during filament extrusion to control the conductivity and dielectric properties of the resulting filament. It may thus be desirable to find a source of carbon-loaded PLA pellets---or pulverized PLA to mix with a powdered additive---to maintain this advantage and combine it with the ease of use of PLA.

The complex relative dielectric functions of the carbon-loaded PLA filament, the carbon-loaded HIPS, plain PLA,\cite{pla} and plain HIPS\cite{hips} plastics were characterized using a pair of filled rectangular waveguide sections [WR28; broad-wall 0.280'', guide height 0.140''] for each material, which each form a Fabry-P\'erot resonator; the waveguide section thicknesses were nominally chosen such that the thinner and thicker sections would constrain the real, $\epsilon_r'$, and imaginary, $\epsilon_r''$, components of the dielectric function, respectively.\cite{wollack2008} To fill the waveguide sections, the plastic samples were heated until soft and then compressed, eliminating air bubbles; excess material was then removed, and the sections were lapped flat. Using a modeling approach described in existing literature,\cite{wollack2008} the dielectric functions were extracted from scattering parameters measured from \SIrange{22}{40}{\giga\hertz} using a vector network analyzer (VNA)\cite{vna} calibrated with Thru-Reflect-Line (TRL) standards, sampled at 801 points. The results of these measurements are summarized in \cref{tab:dielectric}. These measurements target the lower range of the absorber's response, which, in viewing the absorber as a lossy adiabatic structure, has the highest sensitivity to the magnitude of the dielectric function and to the geometry. In practice, the dielectric function for a dilute non-magnetic conductively-loaded thermal plastic is anticipated to be a weak function of frequency\cite{sihvola1999} as has been observed in dielectric mixture systems\cite{wollack2008} analogous to those employed here.

Of the materials tested, the carbon-loaded HIPS is the best candidate for an absorber, since it is an absorptive dielectric that is not too reflective. The carbon-loaded PLA is much more conductive and behaves like a poor metal, making it too reflective for this application. Without carbon loading, the HIPS is a low-loss dielectric, which makes it potentially useful in millimeter wave optics applications, while the PLA is a relatively lossy dielectric, with properties similar to that of nylon.

\begin{table}
\centering
\begin{tabular}{lS[table-format=2.3]S[table-format=2.4]}
\toprule
Material & $\epsilon_r'$ & $\epsilon_r''$ \\
\midrule
Avantra 8130 HIPS & 2.49 & 0.003 \\
Ingeo 4043D PLA & 2.80 & 0.03 \\
PS-715 Conductive HIPS & 7.7 & 2.1 \\
Proto-pasta Conductive PLA & 15. & 15. \\
\bottomrule
\end{tabular}
\caption{Dielectric function measurement results for bulk plastic samples in waveguide section fixtures are presented. The observed relative statistical error of the dielectric function for the average of each set of samples is $<0.01$ for the real component and $<0.12$ for the imaginary component. The systematic error due to calibration errors and sample preparation variation is estimated to be $<2\,\%$ for the real component and $<10\,\%$ for the imaginary component.}
\label{tab:dielectric}
\end{table}

\section{Electromagnetic modeling}

\begin{figure}
\centering
\includegraphics[width=0.75\linewidth]{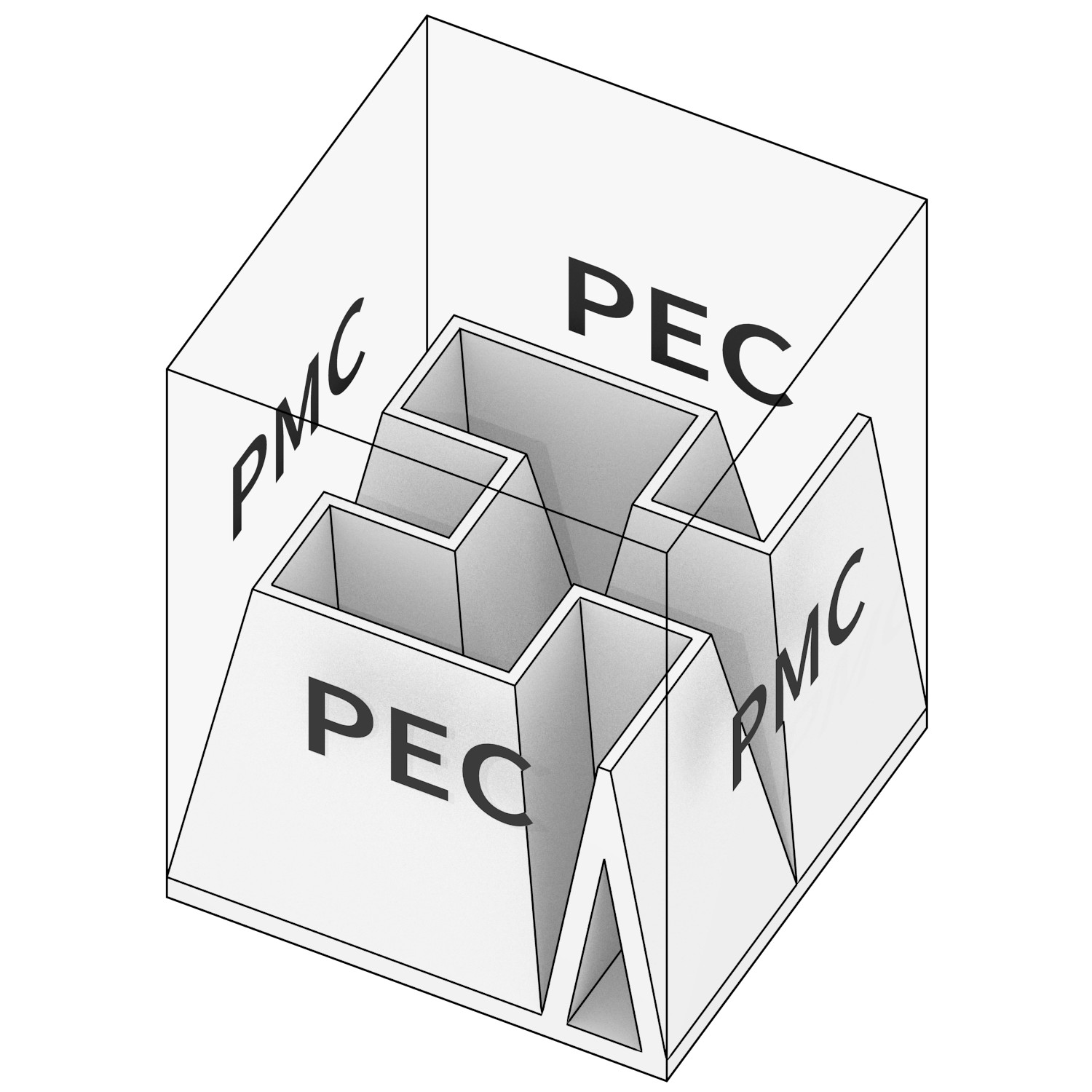}
\caption{The unit cell used in simulations, a second order Hilbert curve, is shown. Periodic boundary conditions were established by using perfect electric conductors (PECs) for the front and back faces of the bounding prism and perfect magnetic conductors (PMCs) for the left and right faces of the bounding prism. The top face of the prism was used as a wave port, and the bottom face was set as a PEC.}
\label{fig:simulation-model}
\end{figure}

A small unit cell of the absorber was modeled in the ANSYS HFSS software package,\cite{hfss} which performs an electromagnetic finite element analysis (FEA). Periodic boundary conditions were implemented using a pair of perfect electric conductor (PEC) planes and a pair of perfect magnetic conductor (PMC) planes around the unit cell. A second order Hilbert curve was used as a unit cell for the absorber, since it has greater symmetry than a first order curve---and thus should have lower residual polarization response---while having reduced simulation complexity when compared to higher order curves. The simulation setup is shown in \cref{fig:simulation-model}. Reflectivity at normal incidence was simulated.

\begin{figure}
\centering
\includegraphics{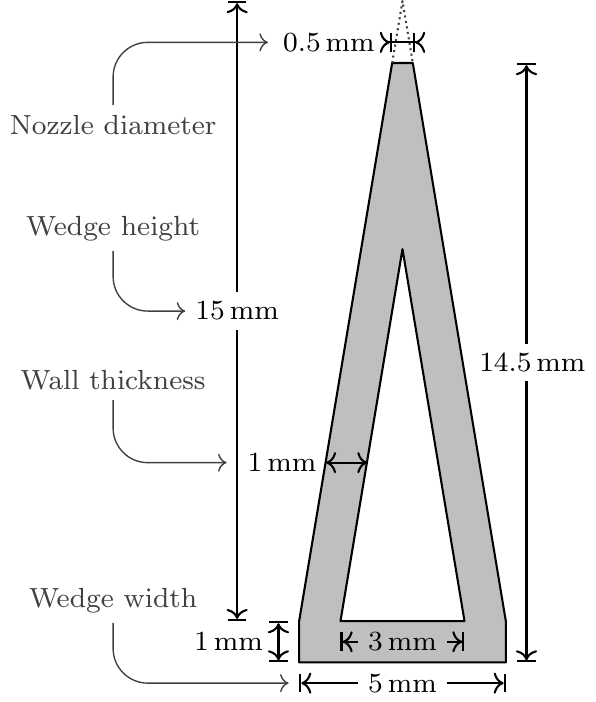}
\caption{A cross section of the absorber wedge is shown, labeled with the dimensions used for the measured prototype, which utilizes a double wall printed with a \SI{0.5}{\milli\meter} nozzle. Due to the nozzle diameter, the wedge tip is truncated.}
\label{fig:cross-section}
\end{figure}

Different wall thicknesses, including a solid cross section, were simulated for the conductive HIPS, as were orthogonal polarizations. Additionally, the conductive HIPS was simulated with varying degrees of conductivity. Possible wall thicknesses are quantized by the diameter of the 3D printer's extrusion nozzle, \SI{0.5}{\milli\meter} in this case. Unless otherwise noted, a double nozzle diameter wall thickness, with a thickness of \SI{1}{\milli\meter} and manufactured with two parallel passes of the printer's extrusion nozzle, and a 3:1 wedge height to wedge width aspect ratio was used (before truncation due to nozzle diameter). A 2:1 aspect ratio was also tried; since this affects  $p/\lambda_g$, it should result in a somewhat different frequency response. Finally, a simulation was performed without wedge tip truncation to evaluate how the geometry performs when it is not limited by manufacturing constraints. To constrain the design space, a fixed height for the absorber was used.

\begin{figure*}
\centering
\includegraphics[width=6in]{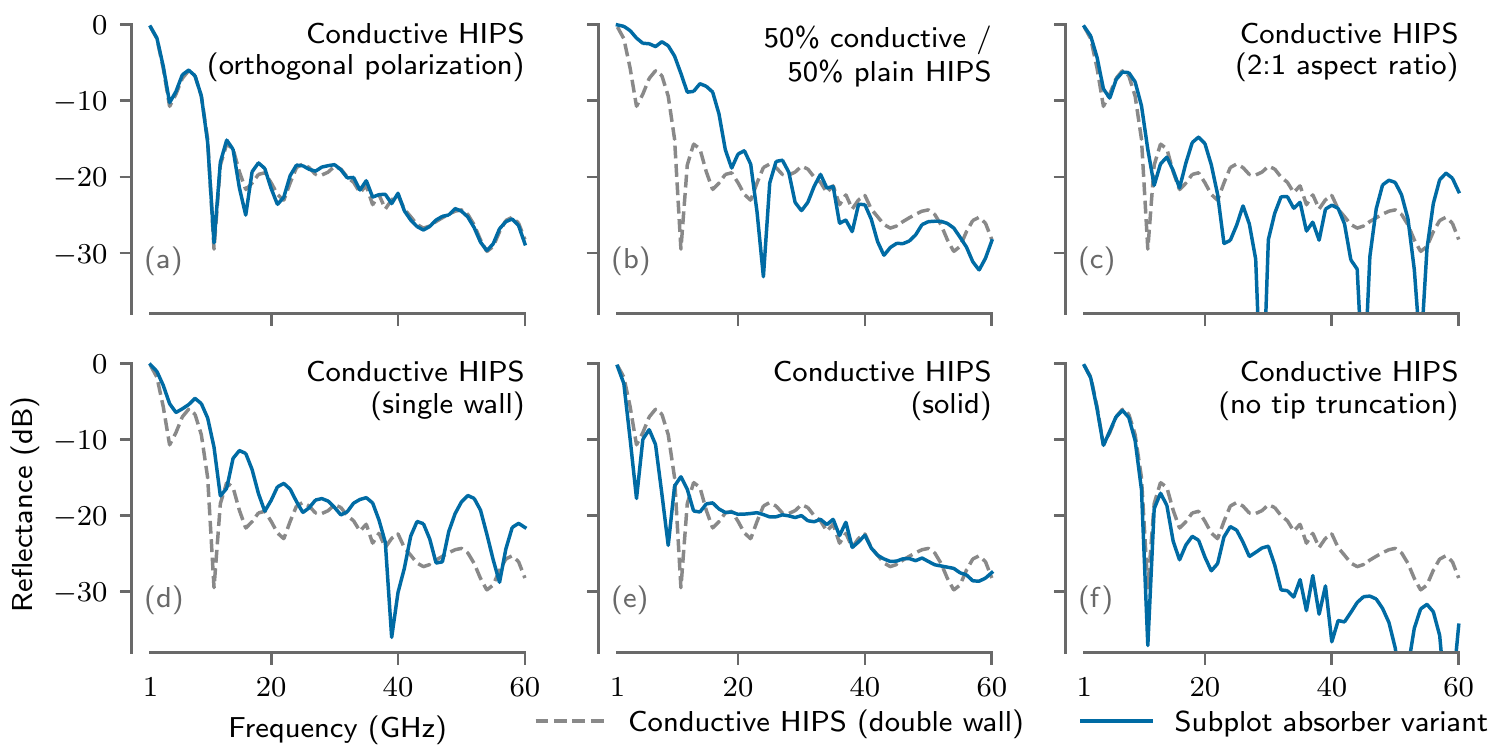}
\caption{The results of FEA reflection simulations of absorber variants (solid) are shown against the baseline simulation of the geometry and material used for the fabricated prototype (dashed). Each simulation was performed from \SIrange{1}{60}{\giga\hertz} at an interval of \SI{1}{\giga\hertz}. The reflectance is shown as a function of frequency for six absorber variants.}
\label{fig:simulation-plot}
\end{figure*}

A cross section of the simulated model, including dimension labels, is shown in \cref{fig:cross-section}; the aspect ratio refers to this wedge cross section, not the unit cell shown in \cref{fig:simulation-model}. The single wall width simulation used a \SI{0.5}{\milli\meter} thick wall, while the triple wall width simulation used a \SI{1.5}{\milli\meter} thick wall. To simulate orthogonal polarization, the PEC and PMC boundaries were switched to the opposite walls from what is shown in \cref{fig:simulation-model}. Simulations were performed from \SIrange{1}{60}{\giga\hertz} at an interval of \SI{1}{\giga\hertz}. Higher frequencies were omitted from simulations to reduce the computation time required; since the reflectivity levels off for higher frequencies, the frequency range simulated is enough to evaluate the relative performance of the simulated absorber variants. The results of these simulations are shown in \cref{fig:simulation-plot}.

The simulation data show good polarization symmetry, which supports the choice of unit cell. The equal mix of conductive HIPS and plain HIPS shows slightly better performance than the conductive HIPS alone, but it was decided that this marginal improvement was not worth the extra effort in filament manufacturing, so a mix was not used for prototyping. The solid cross section simulation results were quite similar to the double wall simulation results, so the double wall was used, as this reduces printing time and material usage without negatively affecting the reflectivity. The single wall and 2:1 aspect ratio simulations show better performance at some frequencies but are worse at others and are more inconsistent across a range of frequencies, so these design variations were not used for the prototype absorber. The similarity of the reflectivities of the geometry and material variants is expected, since adiabatic graded index absorbers are relatively insensitive to their exact geometry and dielectric function.\cite{janaswamy1992,holloway1997}

When the wedge tip was not truncated, reflectivity was significantly decreased, showing that the flatness of the wedge tip that results from manufacturing limitations is a factor that limits performance. If the physical dimensions of the absorber are scaled up, the relative size of the tip truncation will decrease, since the diameter of the nozzle tip---and thus the absolute size of the tip truncation---remains constant. Due to this decrease in the relative size of the tip truncation, scaling up the absorber size should improve the reflectivity beyond what would be expected from the shift due to the change in $p/\lambda_g$. However, this potential increase in performance must be weighed against the increase in absorber thickness.

\section{Fabrication}

To prepare a solid model for an absorber, a script was written to procedurally generate the model to meet specific design parameters using a solid modeling scripting library.\cite{cadquery} With the solid model in hand, G-code instructions for the 3D printer were generated using slicer software.\cite{cura} The solid model generation code, the resulting solid models, and the final G-code have been made available.\cite{zenodo}

\begin{figure}
\centering
\includegraphics[width=\linewidth]{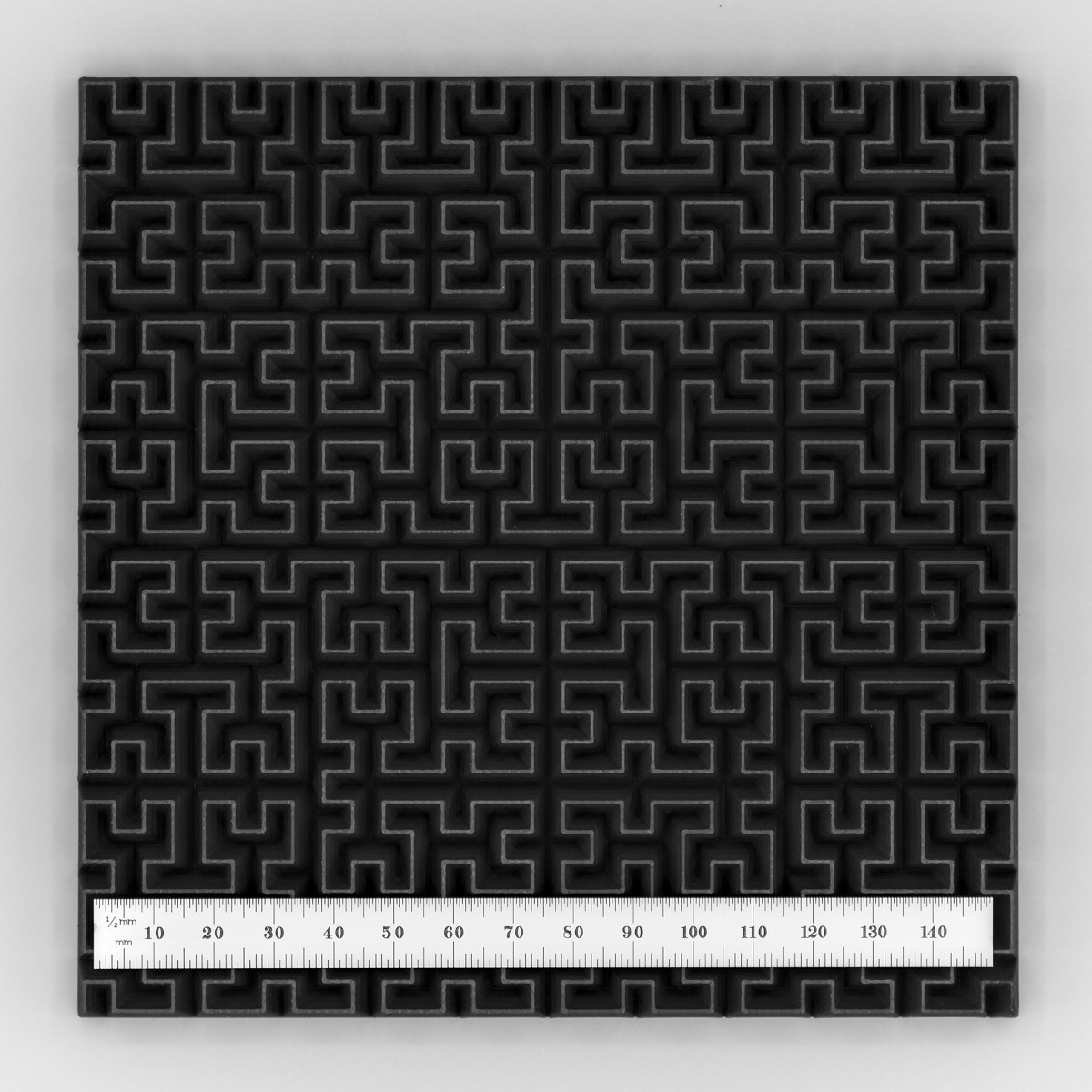}
\caption{The prototype carbon-loaded HIPS Hilbert curve absorber is shown. The footprint of the absorber is a \SI{160}{\milli\meter} square.}
\label{fig:printed-absorber}
\end{figure}

\begin{figure}
\centering
\includegraphics[width=\linewidth]{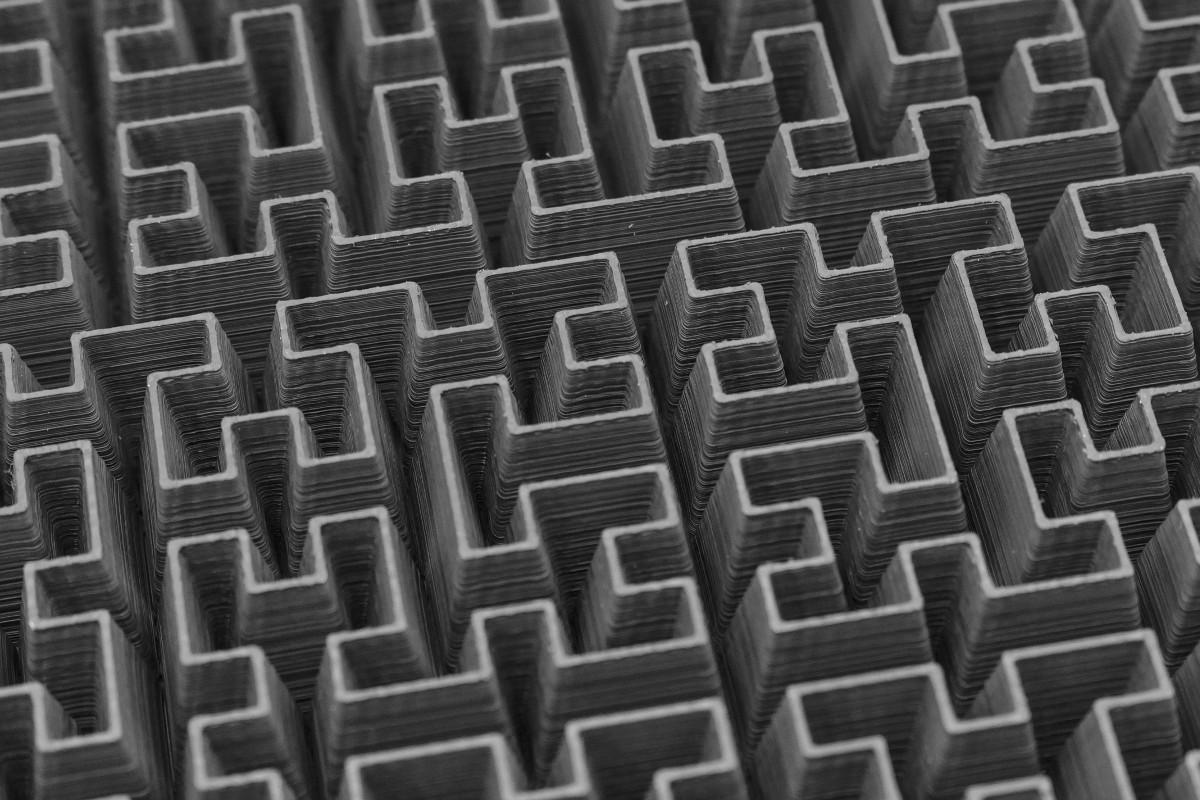}
\caption{A detailed view of the prototype absorber is shown. Layer lines from the 3D printing process are visible.}
\label{fig:printed-absorber-detail}
\end{figure}

This G-code instruction file was then used to print the absorber prototype on a LulzBot TAZ 6 FFF 3D printer\cite{taz6} with a \SI{0.5}{\milli\meter} diameter extrusion nozzle. The tested absorber prototype was manufactured using carbon-loaded HIPS and is a fifth order Hilbert curve with a square footprint of \SI{160}{\milli\meter} by \SI{160}{\milli\meter} and a total height of \SI{14.5}{\milli\meter}; a double wall and 3:1 aspect ratio were used. A cross section of the tested absorber is shown in \cref{fig:cross-section}. The prototype absorber is shown in \cref{fig:printed-absorber}, and a detailed view of the prototype is shown in \cref{fig:printed-absorber-detail}.

\section{Measurement}

\begin{figure}
\centering
\includegraphics[width=3.3in]{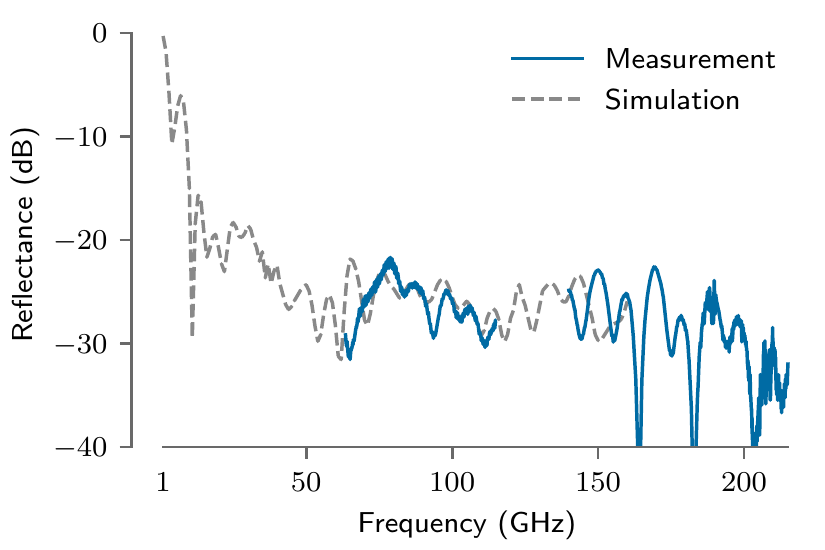}
\caption{Reflectance measurement results (solid) are shown along with reflectance values predicted by electromagnetic simulation (dashed).}
\label{fig:results}
\end{figure}

The fabricated absorber prototype was measured using a VNA coupled to a free space quasi-optical setup. This setup is described in detail in previous literature.\cite{chuss2017} Measurements were taken in two waveguide bands, from \SIrange{63}{115}{\giga\hertz} and from \SIrange{140}{215}{\giga\hertz}; in terms of $p/\lambda_g$, this is 2.9 to 5.3 and 6.5 to 10.0. The results are shown in \cref{fig:results}. While the simulations were performed with a PEC backing behind the absorber, no such conductive backing was used when performing the measurements; however, the power transmitted through the absorber was less than \SI{-50}{\decibel}, so the effect of the backing or lack thereof is insignificant. There is good agreement between the measurements and electromagnetic simulations, providing additional validation of the choice of unit cell used for the simulations. Furthermore, prototypes have been successfully used cryogenically at \SI{60}{\kelvin} for stray light absorption, surviving dozens of thermal cycles, and have been successfully cooled using liquid nitrogen for use as cold loads.

\section{Conclusion}

A broadband graded index absorber was designed around a geometric approximation of the space-filling Hilbert curve and was 3D-printed using FFF. Its reflectance was then measured in the frequency range \SIrange{63}{215}{\giga\hertz} and found to be better than \SI{-20}{\decibel} at normal incidence, which is suitable for the stray light absorption application being considered; evaluation at oblique incidence is left for future work. The use of a space-filling curve overcomes the limitations imposed by FFF while also providing additional mechanical robustness when compared to a traditional tiled pyramid design. With a single 3D printer, one absorber can be fabricated per day, but this production rate can be easily scaled up by parallelizing fabrication across multiple 3D printers. To extend the concept from stray light absorbers to cryogenic calibration targets, the reflectivity needs to be further reduced, and the thermal gradient caused by plastic's poor thermal conductivity needs to be addressed, such as through the use of a thermalizing core, which could be made by adding a metal insert into holes left in the bottom of the 3D print or by filling the currently hollow interior with an epoxy designed to have higher thermal conductivity.

The absorber could be further refined by using a smaller extrusion nozzle on the 3D printer, allowing for a sharper peak to the wedge, which should improve performance. Since additive manufacturing allows for rapid prototyping, multiple wedges profiles could be easily tried and their performances compared. The use of a multi-material FFF printer could allow for different amounts of carbon loading to be used for the outer and inner walls of the absorber to decrease reflection or to potentially create a gradient in the carbon loading, something that would be difficult to accomplish with more traditional manufacturing techniques. Moving away from FFF, a selective laser sintering (SLS) [stereolithography (SLA)] 3D printer could be used with carbon powder or stainless steel powder mixed in with the nylon powder [resin] to further extend the ease of customization of the absorbers, including for covering curved surfaces, by utilizing a more traditional pyramidal absorber structure. The fine resolution of SLA 3D printers is also well suited for the creation of extremely low reflectivity calibration targets.

\begin{acknowledgments}
The authors would like to thank Marco Sagliocca for his help in performing the quasi-optical measurements. This work was supported by a Space\makeatletter @\makeatother{}Hopkins seed grant. J. Appel, C. Bennett, J. Eimer, T. Marriage, and M. Petroff are supported via National Science Foundation Division of Astronomical Sciences grants for the Cosmology Large Angular Scale Surveyor (CLASS), grant numbers 0959349, 1429236, 1636634, and 1654494.
\end{acknowledgments}

\bibliography{paper.bib}

\end{document}